\documentclass[twocolumn,amsmath,amssymb,prl,footinbib,floatfix,superscriptaddress]{revtex4}

\usepackage{graphicx}
\usepackage{amsmath}
\usepackage{amsfonts}
\usepackage{mathrsfs}
\usepackage{color}
\usepackage{slashed}
\usepackage{stackrel}

\def\<{\langle}
\def\>{\rangle}

\def\beq{\begin{equation}}
\def\eeq{\end{equation}}
\def\barray{\begin{eqnarray}}
\def\earray{\end{eqnarray}}

\newcommand{\ket}[1] {\vert{#1}\rangle}

\newcommand{\be}{\begin{equation}}
\newcommand{\ee}{\end{equation}}
\def\ba{\begin{eqnarray}}
\def\ea{\end{eqnarray}}

\def\tr{\bigtriangleup} 
\def\tri{_\bullet \mkern -  \thickmuskip   \stackrel{\bullet}{\bigtriangleup} _\bullet}

\font\numbers=cmss12
\font\upright=cmu10 scaled\magstep1
\def\stroke{\vrule height8pt width0.4pt depth-0.1pt}
\def\topfleck{\vrule height8pt width0.5pt depth-5.9pt}
\def\botfleck{\vrule height2pt width0.5pt depth0.1pt}
\def\Zmath{\vcenter{\hbox{\numbers\rlap{\rlap{Z}\kern
0.8pt\topfleck}\kern 2.2pt
                   \rlap Z\kern 6pt\botfleck\kern 1pt}}}
\def\Qmath{\vcenter{\hbox{\upright\rlap{\rlap{Q}\kern
                   3.8pt\stroke}\phantom{Q}}}}
\def\Nmath{\vcenter{\hbox{\upright\rlap{I}\kern 1.7pt N}}}
\def\Rmath{\vcenter{\hbox{\upright\rlap{I}\kern 1.5pt R}}}
\def\Pmath{\vcenter{\hbox{\upright\rlap{I}\kern 1.5pt P}}}

\begin{document}

\date{\today}
\title{Holographic codes}
\author{Jos\'e I. Latorre$^1$ and Germ\'an  Sierra$^2$ \vspace{0.2cm} \\  
${}^1$ Departament d'Estructura i Constituents de la Mat\`eria, Universitat de Barcelona, 
Barcelona, Spain,   \\ 
Centre for Quantum Technologies, National University of Singapore, Singapore.\\
 ${}^2$  Instituto de F\'isica Te\'orica UAM/CSIC, Universidad Aut\'onoma de Madrid, Cantoblanco, 
 Madrid, Spain.}

\begin{abstract}
There exists  a remarkable four-qutrit state  
that carries absolute maximal entanglement in all its partitions.  Employing 
this state, we construct a tensor 
network that delivers a holographic many body state,
the H-code,  where the physical properties of the boundary determine those of the bulk. 
This H-code is  made of an even  superposition of states whose relative Hamming distances are exponentially
large with the size of the boundary. This property makes  H-codes  natural states for a quantum memory.
H-codes  exist on tori of definite sizes and get classified in three different sectors 
characterized  by  the sum of their qutrits on cycles wrapped through the boundaries of the system.  
We construct a parent Hamiltonian for the H-code
which is highly non local and finally we compute the topological entanglement entropy of the H-code. 
\end{abstract}

\maketitle

We introduce the concept of a holographic code as a balanced superposition of quantum states 
for which boundary and bulk degrees of freedom are strictly related 
\beq
|H\rangle = \frac{1}{\sqrt {d^n}}\sum_{i=1}^{d^n} |i\rangle_{\rm boundary} |\phi_i\rangle_{\rm bulk} ,
\label{pab}
\eeq
where  the states $\ket{i}$ form a product basis of the Hilbert space of the boundary  made of $n$ degrees of freedom of dimension $d$,
and $\ket{\phi_i}$ are orthogonal product states in the bulk, $\langle\phi_i | \phi_j\rangle=\delta_{ij}$.
Let us emphasize that all $\ket{i}$ and $\ket{\phi_i}$ are product states, at variance with the structure of
the standard Schmidt decomposition where such states are not necessarily product states. 
We shall also construct substates of a holographic code, that
will be characterized by observables that test the topology of the state.
A good property for a holographic code will later be characterized by large Hamming distances among all elements $\ket{\phi_i}$, so that distinguishability of each individual element is maximized. 

It follows from the basic definition in Eq.(\ref{pab})  that the scaling of entanglement entropy in orthogonal holographic codes is bounded by an area law. This upper bound emerges trivially from the fact  that the total amount of superpositions for any partition of the system is bounded by 
$d^n$.
This key feature is guarantee by the balanced superposition of states and the
fact that the bulk states are all product states.

\bigskip
{\sl Tensor network construction of an orthogonal  holographic code.}--
Let us consider a quantum system made of qutrits on an infinite triangular lattice in 2D.
We shall now define a tensor network that gives rise to a class of orthogonal 
holographic states.

\begin{figure}[ht]
\centering
\includegraphics[width=0.25\textwidth]{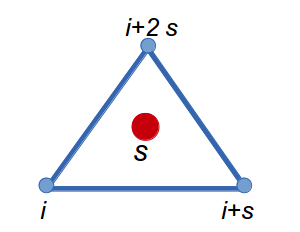}
\caption{Basic simplex underlying the orthogonal holographic network. The relation
between the physical index $s$ and the ancillae is dictated by the absolute maximally
entangled state given in Eq.(\ref{simplexstate}). 
 } 
\label{simplex} 
\end{figure}

The basic idea is to construct a network of triangular simplices on triangles
pointing up,  where ancillary
qutrits live. Then a physical qutrit index is dictated by the values of the
underlying ancillary indices. The construction is illustrated in Fig. \ref{simplex}.
The specific assignment for each physical qutrit is made as follows
\be
  |\psi\rangle = \sum_{i,s=-1,0,1} |s\rangle_{\rm physical} |i, i+s,i+2 s\rangle_{\rm ancillae}
\label{simplexstate}
\ee
where the index $s$ corresponds to the physical  index, 
and all qutrit indices live in $Z_3$ and are to be considered mod(3). 
This construction is tantamount to set up a PEPS-like tensor network
based on the translational invariant tensor $A^s_{i,j,k}$ which is
1 if $j=i+s$ and $k=i+2s$ and 0 otherwise.
The rational for this construction
is that such a state is an example of absolute maximal entanglement \cite{HCRLL12}. 

Let us note that the existence of absolute maximally entangled states is non-trivial \cite{GW11}. For instance,
there are no four-qubit states that have such a property. It is possible to construct absolute 
maximally entangled states related to Reed-Solomon codes, which only appear for certain local
dimensions and number of local Hilbert spaces \cite{HCRLL12}. 
The four-qutrit state Eq.(\ref{simplexstate}) 
we use in our construction
is the first non-trivial maximally entangled state for four local degrees of freedom, that is, the
 first fully entangled generalization of EPR states to the four-body case.

The first relevant property of the state Eq.(\ref{simplexstate}) we have taken as the underlying structure of the
tensor network is that it provides a mapping of a 2-qutrit basis onto a related but different 2-qutrit basis.
Indeed, the indices $i$ and $s$ span the natural basis for two qutrits and
$i+s$ and $i+2 s$ produce a second basis. Therefore, given the value of two qutrits,
the other two are fixed, at variance with traditional Projected Entangled Pairs States (PEPS) where two indices do not
fully determine the other two \cite{VC04,CV09}. Therefore, in our construction, 
fixing a first raw of physical indices and a unique
ancilla, the rest of the network, including all physical indices, is fixed. Furthermore,
this construction is also valid as seen from the diagonal directions, since the 
property of absolute maximal entanglement guarantees that it is always possible to 
view the network assignment as a mapping between basis, no matter which partition of
the four indices is made.

\begin{figure}[h]
\centering
\includegraphics[width=0.45\textwidth]{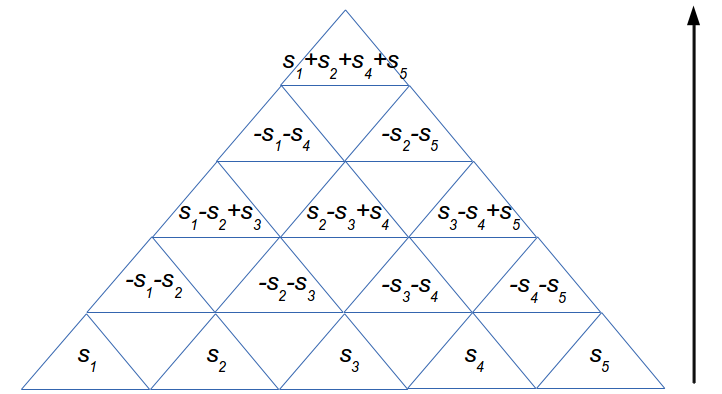}
\caption{Network construction of the H-code based on the absolute maximally entangled simplex.
Physical indices are shown as their label in the center of ancillary triangles. The net
effect of the ancillae is to make  the sum of the values of physical qutrits
forming and up triangles to be zero (mod 3). All qutrits in the bulk are
defined following the holographic direction shown in the right side. This
holographic direction can be chosen differently.
 } 
\label{network} 
\end{figure}

The second relevant property emerging from the absolute maximally entangled state of Eq.(\ref{simplexstate}) is that
the second row in the network of physical indices is completely determined in a simple way. It is easy to check that
if two physical indices on a first raw are taken to be $s_1$ and $s_2$, then they force the physical
index in the second raw to be $-s_1-s_2$ (see Eq.(\ref{neu})). This is tantamount to define the  upper physical
index by imposing that the total sum of indices is zero, where we always work mod 3 because
of the qutrit nature of the local Hilbert spaces.
This systematic enforcement of the values of all
physical indices proceeds as more rows are added to the network as shown in Fig. \ref{network},
so that a perfect holographic state emerges. All the elements in the bulk are determined
from the qutrits in the boundary.

At this point of the construction of the holographic state it is possible to 
dispose of the underlying ancillary structure. The net effect of the ancillary states
was to obtain a state based on the simple rule that every triangle of physical indices
must add up to zero mod 3, namely
\beq
s_1 + s_2 + s_3 = 0 \; ({ \rm mod }  \;  3)
\label{neu}
\eeq
 We shall call this property the {\sl neutralization rule}.
Once this neutralization rule is deduced, it suffices to produce all
the state as if we were dealing with a cellular automaton operation.  
Given the first row of $n$ qutrits $s_1\ldots s_n$, the complete state is represented by 
\be
\ket{H_{s_1,\ldots,s_n}}=\ket{s_1,\ldots,s_n}_{\rm boundary}\ket{f(s_1,\ldots,s_n)}_{bulk}
\label{H}
\ee
where $f(s_1,\ldots ,s_n)$ delivers the values of each qutrit in the bulk using repeatedly
the neutralization rule.  It is clear
that this neutralization rule acts as a cellular automata defining a row at a time. Every physical
qutrit is defined by its predecessors that form a sort
of backwards light-cone.

Let us note that all states based on homogeneous tensor networks, such as translational
invariant PEPS \cite{VC04}-\cite{O14}, are
holographic, in the sense that the value of the boundary ancillae determine a
superposition state in
the bulk. This also applies to the  H-state since it is also determined by the simplex
structure of the ancillae.
However, the difference is that standard tensor networks
are such that orthogonal states  on the boundary produce non-orthogonal 
states in the bulk, while in our case orthogonality in the bulk is preserved. 
Moreover, we  shortly prove that the neutralization rule enforces a topological
structure absent in usual PEPS.

From now on we shall refer to the superposition state 
\be
  \ket{H}=\frac{1}{\sqrt{3^n}}\sum_{s_1,\ldots,s_n=-1,0,1} \ket{s_1,\ldots,s_n} \ket{f(s_1,\ldots,s_n)}
  \label{H-code}
\ee
as a H-code, standing for our orthogonal holographic code. Note that $n$ is the length  of the boundary
and the bulk can extend to a number of rows $m$ that depends on
the topology of the system. 

\bigskip
{\sl H-code on a torus.}-- 
It is not obvious that the H-code can be defined on lattices with non trivial topologies such  as the  torus.
The reason is that the neutralization rule defines every layer of the state
and may be  not allow for periodic boundary conditions.

The solution to this riddle is somewhat surprising. We shall now see
that H-codes can be defined on a torus of size $n\times m$ only for certain values of
$n$ and $m$. In particular, there is a H-code for every torus $n=m=3^k$.
To obtain this result, let us consider Fig. \ref{network}. Periodicity
will be allowed if a consistent identification of spins along the rows and columns is possible.
Let us focus on the first qutrit of each row and notice
that in the fourth  row we find it to be $-s_1-s_4$. This allows for the assignment
$s_1=s_4$ that brings $-s_1-s_4=-2 s_1=s_1$, thus providing exact periodicity both
in the diagonal and the row for the case $n=m=3$.
%

For larger tori, a scale symmetry emerges. It is easy to see that the
tenth row first qutrit reads $-s_1-s_{10}$. Then, the identification
$s_1=s_{10}$ yields again perfect periodicity for $n=m=9$ since
$-s_1-s_{10}=-2 s_1=s_1$. This pattern
linking the horizontal period with the diagonal one
is repeated for every $n=m=3^k$ with $k \geq 1$ (see Supplementary material). 
It is possible to find other periodic H-codes in the holographic dimension
depending on the values of $(n,m)$. For instance, we have found  valid H-codes
on tori $(n,m)=(5,40),(7,182),(11,121),\ldots$. For other cases, only a subsector of the boundary
Hilbert space can yield periodic sates. In the following, we shall stick
to the squared tori with size $n=m=3^k$ with perfect periodicity.

There is a further relevant property found on H-codes on a torus. It is easy to verify
that for sizes $n=m=3^k$, the product of the qutrit values on every horizontal
line as well as every diagonal line is equal to
\be
  Q=\prod_{i=1,\ldots,3^k}  z_i   \ ,
  \label{constant}
\ee 
where each $z_i = q^{1+s_i}$ can take values 1, $q$ or $q^{2}$, corresponding to
the spins $s_i=-1,0,1$,  with $q=e^ {2 \pi i/3}$.
As a consequence, a sub-structure of states within the H-code into three
categories emerges, each one labelled by a different value of $Q$.
Such a property hints at the possibility of using topological H-code as a quantum memory.

\bigskip
{\sl Hamming distance for the H-code.}-- 
It is interesting to see how a H-code state can store information. 
Given the cellular automata character of the neutralization rule,
any change of a given qutrit propagates a modification of the state
on a sort of light-cone which is contained within the diagonals emerging
from that point upwards. 
This propagation of changes is necessary to preserve
 the neutralization rule Eq.(\ref{neu}). 
If a single qutrit is changed in the boundary, then every row
in the bulk needs to be changed.

It is possible to verify that any two states in the H-code on a torus
of size $3^k$, differ at least by $6^k$ elements. That is, the minimum
Hamming distance between any pair of elements in the code
is $dist_H=6^k$. This property makes the elements of the code easily 
distinguishable, which provides a way to code information in 
a very redundant manner. As an example, for a $9\times 9$ code, there are
a total of $3^9=19683$ states, as defined in the 9 qutrit boundary.
Then, in the large Hilbert space made of 81 qutrits, all the elements
in the superposition differ at least by a
Hamming distance of $dist_H=36$. 

The relevant point is that  H-codes provides exponential 
distinguishability as the size of the torus increases.
Furthermore, non-orthogonal holographic codes such as PEPS do
not allow for easy distinguishability, since bulk configurations 
are not orthogonal.

\bigskip
{\sl Construction of a H-code.}-- 
The simplest way to generate a H-code is to guarantee a superposition of
elements in the boundary and then set the rest of elements using a
cellular automata strategy. 
We first encode the previous index $s$  as an exponent,
so as to implement
the qutrit nature of the local physical system. That is $e^{2 \pi i s/3}$
gives rise to $1$ for $s=0$, $q$ for $s=1$ and $q^{2}$ for $s=2$.

 We choose as boundary Hamiltonian
\be 
  H_{\rm boundary}= -  \sum_{i=1}^{n} \sum_{a+b=0} X_i^a X_{i+1}^b
\label{Hboundary}
\ee
where the sum over $i$ runs through the $n$ qutrits in the boundary and
\be
  X=
  \left( \begin{array}{ccc}
0 & 0 & 1 \\
1 & 0 & 0 \\
0 & 1 & 0 \end{array} \right)\ ,
\label{X}
\ee
acts as a raising operator (note that $X^2=X^{-1})$. 
 The idea behind
this choice of boundary Hamiltonian is that it performs a generalized symmetrization 
of every pair of qutrits while preserving their sum, that is, it maps
(00,12,21) onto (00+12+21), (01,22,10) onto (01+22+10) and (02,11,20) onto (02+11+20).
Note that the symmetrization
implied by $H_{\rm boundary}$ makes it relevant to take open or periodic boundary
conditions.

\begin{figure}[ht]
\centering
\includegraphics[width=0.40\textwidth]{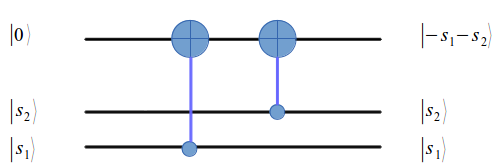}
\caption{
The action of each two body gates corresponds to $\ket{a}\ket{b}\to 
\ket{a}\ket{b a^2}$ where $a,b=1,q,q^2$, with $q=e^{2\pi i/3}$.
This is identical to the neutralization rule expressed in terms of $s=0,1,2$ which
appears in the exponent $a,b=e^{2 \pi i s/3}$.
 } 
\label{unitary} 
\end{figure}

We can now propagate this superposition into a bulk using the following
cellular automata strategy. For each layer we act on contiguous qutrits using
the unitary operator
\be
U \ket{s_1}\ket{s_2} \ket{0} = \ket{s_1}\ket{s_2}\ket{-s_1-s_2}\ .
\ee
which can be created using the circuit in Fig. \ref{unitary}.
\bigskip

The construction we have provided, based on a boundary Hamiltonian
and a cellular automata,  has  three distinct 
ground states
\be
  \ket{H}_S =\sum_{s_1+\ldots+s_{n}=S}\ket{s_1,\ldots,s_n}\ket{f(s_1,\ldots,s_{n})}
   \ ,
\label{GS}
\ee
labeled by the sum of the qutrits on any row or diagonal  $S=0,1,2$,
which corresponds to $Z=e^ {\frac{2 \pi i}{3} S}$.
This shows that the H-code is able
to codify one qutrit state through properties related to observables
spanning through  boundaries of the system. The fact  that $S$ takes
the same value on every row and every diagonal and
the large Hamming distance among all superpositions in the H-code 
suggests that the coding
of states is robust against some local fluctuations.

\bigskip
{\sl Parent Hamiltonian.}--
Tensor network states have the property of being the ground states 
of the so called parent Hamiltonians. They do not have to be  unique. 
For the H-code there is a huge freedom (see  Suplementary material). 
The simplest case is  made of two terms 
\beq 
  H=H_Z + H_X 
  \label{pa0}.
\eeq
The first term reads
\beq
H_Z =  \sum_{i,j,k \in \tr  }  \left( 2 - Z_i Z_j Z_k -  (Z_i Z_j Z_k)^2 \right) \, , 
\label{pa1}
\eeq 
where 
\be
  Z=
  \left( \begin{array}{ccc}
1 & 0 & 0 \\
0 & q & 0 \\
1 & 0 & q^2 \end{array} \right)\ , \quad q= e^{ 2 \pi i /3} \, . 
\label{Z}
\ee
The sum runs over all  the spins located at the corners of the  up triangles  
of the lattice.  One can easily verify that $H_Z$ vanishes if and only if
$s_i + s_j + s_k =0 \; ({\rm mod} \;  3)$. 
In the remaining cases $H_Z=3$  for each violation of the neutrality rule Eq.(\ref{neu}).
The Hamiltonian (\ref{pa1}) is highly degenerate since any configuration satisfying 
Eq.(\ref{neu})  will be a ground state. To break this degeneracy
we consider the operator  (see Suplementary material) 
\beq
H_X  = - \sum_{n= \pm1}  \sum_{p =0, \pm 1  }  \prod_{a,b=1, \dots, 3^k}  X_{a,b}^{n (a-b + p)}  \, ,  
\label{pa111}
\eeq 
where $X_{a,b}$ denotes the operator (\ref{X}) acting on the site $(a,b)$ 
of the $3^k \times 3^k$ torus. $H_X$ contains a pair of operators $X$ and $X^{-1}$
per each  up triangle, and an equal number of $X's$ and $X^{-1} \,  's$ per row or column.
These properties imply that $H_X$ commutes with $H_Z$ and $Q$, hence the ground states
of $H$ satisfy the neutrality rule, that  minimizes $H_Z$,  and have a definite value of $Q$.
In these subspace of states, $H_X$ is a non positive matrix and then the
Perron-Frobenius theorem yields  that the  ground state is  
the H-code  Eq.(\ref{GS}) with $Q= e^{ 2 \pi i S}$. $H_X$ is a highly non local operator
that contains  products of $2 \times 3^{2 k -1}$ matrices $X^{\pm 1}$. This non locality 
is due to the exponential increase of the  Hamming distance that grows as  $6^k$. 
One may speculate that non local Hamiltonians such as (\ref{pa111}) may arise
from integrating some gauge degree of freedom. 

\bigskip

{\sl Entanglement entropy and topological entropy.}--
Entanglement entropy remains a natural way to quantify the
amount of quantum correlations in a given state \cite{KP,LW}. It turns out
that an exact computation of the entanglement entropy for the H-code
is easily done. Let us first consider the simplest situation
with three qutrits forming a triangle as in Fig. \ref{simplex}, and let us  call them $A$, $B$ and $C$. 
Qutrits $A$ and $B$ are free to take any value, but qutrit $C$ is
dictated by the neutralization rule, since $A$ and $B$ are its
backward light-cone. Then
\begin{equation}
S_{ABC}=log_3 \,  3^2=2\ ,
\end{equation}
where we have decided to measure 
entropies using base 3, given the qutrit structure of the H-code. 
Similarly, for any close region $A$ made of $m$ qutrits, we must
count how many of them are free {\sl vs.} those who are dictated
by the neutralization rule. So, if $n_{ind}$ independent qutrits appear in
the boundary, the entropy will be
\begin{equation}
S_A=n_{ind} .
\end{equation}

This reasoning is now sufficient to compute the topological entropy as
follows. Consider again the case of a triangle made with three qutrits
$A$, $B$ and $C$, and let us call the rest of the system $D$. Then
the topological entropy will be given by \cite{KP,LW}
\begin{eqnarray}
S_{\rm top}&= & S_{ABC}-S_{AB}-S_{AC}-S_{BC}+S_A+S_B+S_C \nonumber 
\\
& = & 2-2-2-2+1+1+1=-1 .
\end{eqnarray}
This extends to any configuration of contiguous $A$, $B$ and $C$
since it is easy to show that adding one qutrit at a time in any position
around a given configuration preserves the value of $S_{\rm top}=-1$.

The three possible states of the H-code on a torus, that
is $\ket{H_S}$, cannot be distinguished using bulk observables.
The reduced density matrix of any subset $A$ made out of $m$ bulk qutrits, with $m < 3^k$ is simply $\rho_A=\frac{1}{3^m} I_{3^m}$, that is $\rho_A$ is fully disordered.
In the case that the number of qutrits appearing in $A$ is larger than
the size of the boundary of the torus, there are not sufficient degrees of freedom
to achieved maximum entropy. Then, an area law appears for the entropy

\bigskip
{\sl Conclusions.}--
 We have introduced the concept and explicit construction of a holographic code which is
characterized by an exact mapping of boundary  states onto bulk  ones. This mapping
is achieved by a neutralization rule which is related to the ground state of
a two-body nearest neighbour Hamiltonian. The construction of a symmetric state
on the boundary produces a H-code on the bulk of a torus which can encode a qutrit 
through its topological properties. 

In general, an H-code can be viewed as a compressor of bulk states that uses
a basis of elements whose Hamming distances are large. Alternatively, an 
H-code makes redundant the information of its boundary by disseminating it
through the bulk of the system.

\bigskip
{\sl Acknowledgements.}-- We thank D. Cabra, J. I. Cirac, S. Iblisdir, J. Pachos, N. Schuch and F. Verstraete 
for their comments. We acknowledge financial support from FIS2013-41757-P,  FIS-2012-33642, QUITEMAD, and the Severo
Ochoa Programme under grant SEV-2012-0249.

\section*{Suplementary Material}

\bigskip
{\sl H-Codes on the torus}--

Let us consider a torus  of size $(n,m)$.  Using 
the neutralization rule   Eq. (\ref{neu}), the map from row-to-row states
can be represented as 

\beq 
\left( \begin{array}{c}
s'_1 \\
s'_2 \\
\vdots \\
s'_n \\
\end{array}
\right) = 
\left( \begin{array}{rrrrr}
-1 & -1 & 0 &  \dots & 0 \\
0 & -1 & -1&  \dots & 0 \\
\vdots & \vdots & \vdots & \vdots & \vdots \\
-1 & 0 & 0 & 0 & -1 \\
\end{array}
\right)  \left( \begin{array}{c}
s_1 \\
s_2 \\
\vdots \\
s_n \\
\end{array}
\right), \; ({\rm mod } \,  3) \, , 
\label{c1}
\eeq
where $s_i=0,1,2$ (mod 3). 
The $n \times n$ transfer matrix $T_n$ defined in this way  reads

\beq
T_n  = - ( {\bf 1} + U_n),
\label{c2}
\eeq
where ${\bf 1}$ is the $n^{\rm th}$-dimensional identity matrix  and 
\beq 
U_n =  
\left( \begin{array}{rrrrr}
0 & 1 & 0 &  \dots & 0 \\
0 & 0 & 1&  \dots & 0 \\
\vdots & \vdots & \vdots & \vdots & \vdots \\
1 & 0 & 0 & 0 & 0 \\
\end{array}
\right)   \, , 
\label{c3}
\eeq
satisfies

\beq
U_n^n = {\bf 1} \, . 
\label{c4}
\eeq
A  $H$-code on the torus $(n,m)$ is possible   if after $m$ iterations  of $T_n$
any  state on the first row returns to itself.  This  situation  is guaranteed if and only if  
\beq
T_n^m = {\bf 1}   \quad  ({\rm mod } \,  3) \, . 
\label{c5}
\eeq
Let us show that this condition holds for  $n=m=3^k$. Using  Eq. (\ref{c2}) 
\barray 
T^n_n & = &  (-1)^n ({ \bf 1} + U_n)^n = 
\, -  \sum_{r=0}^n 
\left( 
\begin{array}{c}
n \\
r \\
\end{array}
\right) 
U^r_n \label{c6}  \\
& = & - ( { \bf 1} + U_n^n) \, ({\rm mod} \, 3)  =  { \bf 1}  \, ({\rm mod} \, 3) \, , 
\nonumber 
\earray 
where we have used Eq.(\ref{c4}) and  the fact that $\left( 
\begin{array}{c}
n \\
r \\
\end{array}
\right) $ is divisible by 3, for $r=1, \dots, n-1$ with $n= 3^k$. 
This example shows that  finding  consistent $H$-codes
is an interesting  problem in modular arithmetic.

\bigskip
{\sl Injectivity and parent Hamiltonian}--

Let us represent the PEPS-like tensor $A^s_{i,j,k}$  as 
\beq
\bigtriangleup \quad  = \quad   
\tri
\label{h1}
\eeq 
where $\bullet$ indicates the ancilla  indices $i,j,k$ and the   triangle is associated to the   spin $s$. 
On the plane one  can construct two generic types of networks labeled by an integer $k \geq 1 $.
The networks for $k =  1$ are 
\beq  
\begin{array}{ccccc}
 & &  \tri  & & \\
&\tri  & & \tri  & \\
\tri  &  & \tri & &  \tri  \\
\end{array} 
\label{h2}
\eeq
and 
\beq
\begin{array}{ccccc}
\tri  &  & \tri & &  \tri  \\
&\tri  & & \tri  & \\
 & &  \tri  & & \\
\end{array} 
\label{h2b}
\eeq
and their type will be denoted  $\bigtriangleup$  and $\bigtriangledown$ respectively. 
Both  networks 
have $3 k$ triangles aligned on the edges, and a total of  $3 k ( 3 k +1)/2$ triangles, that is spins. 
In the  networks of type $\bigtriangledown$  all the internal
ancillas are contracted in groups of three  using the GHZ state, while in the  $\bigtriangleup$ networks the ancillas
on the edges are contracted using a Bell state.   These networks differ in the fact that  for the   $\bigtriangleup$-type 
the values of the spins on a single edge determine holographically the remaining spins, while 
for  the $\bigtriangledown$-type   one needs  the values of the spins on two edges. 

Let us study the injectivity properties  of these tensor networks. 
 PEPS   is injective if the map constructed with a tensor network   between the ancilla indices and the physical indices
is injective \cite{DP}. 
Here the ancillas refer to the ones that are left uncontracted or open. 
Recall that a  linear map $T: V \rightarrow W$ is injective if the kernel of $T$ is empty. 

To analyze this property  we choose the $\bigtriangledown$ networks  
whose  internal ancillas  are all contracted while the external ones 
are open.  There is a total of $9k$ outgoing  ancillas, but they are 
not all independent: fixing $6 k$ ancillas on two boundaries determine the values
of the remaing outer ancillas and all the spins. This implies that the map from the
outgoing  ancillas and the spins has a non trivial kernel, hence  the map is not injective.  
The same happens for the networks of type  $\bigtriangleup$. For MPS the injectivity property
guarantees the uniqueness of the ground state of the  
parent Hamiltonian  but not for PEPS \cite{DP}. Nevertheless we construct below a parent Hamiltonian
whose ground state on the torus  has only the degeneracy due to the charge $Q$
of Eq.(\ref{constant}).

First notice that the  neutralization rule  Eq. (\ref{neu})  is satisfied by all the 
ground states of the  Hamiltonian  (recall Eq.(\ref{pa1}) )
\beq
H_Z = \sum_{i,j,k \in \tr  }  \left( 2 - Z_i Z_j Z_k -  (Z_i Z_j Z_k)^2 \right)  \, , 
\label{ppa1}
\eeq 
 where  $Z_i$ is defined in Eq.(\ref{Z}). 
The proof is as follows. The 
solutions of Eq.(\ref{neu}) are: 
1)   $s_1 = s_2 = s_3$ that corresponds to $z_1 = z_2= z_3$ equal to $1, q, q^2$
 in which case $z_1 z_2 z_3 =1$ and so $H_z =0$;  and 2)  $s_1=1, s_2 = q, s_3 = q^2$
 (and permutations) that also leads to $z_1 z_2 z_3 =1$ and $H_Z=0$. In the remaining cases where (\ref{neu}) is not satisfied
one finds  that $H_Z=3$ for each frustrated triangle. 
 
 Let us construct an operator that mixes the states satisfying  Eq.(\ref{neu}). First we consider
 the state on the torus   with $3 \times 3$ sites 
 \beq
 \begin{array}{ccccccccc}
        &           &     &   1   &      &  2   &      &  3   &   \\ 
        &           &     &   \tr  &      & \tr   &      & \tr   &   \\ 
        &           & 7  &       &   8 &      &  9  &      &  7 \\ 
        &           & \tr &       &  \tr &      & \tr  &      &     \\ 
        &    4     &     & 5    &      & 6   &      &  4  &   \\ 
        &   \tr     &     & \tr   &      & \tr  &      &      &   \\ 
 1     &           & 2  &       & 3   &      & 1   &      &   \\                 
\end{array}
\label{ttt}
\eeq
$i=1, \dots, 9$ labels  the spins $s_i$. Eq.(\ref{neu}) is satisfied 
on each triangle,  e.g. $s_1+ s_2 + s_4 = 0 \;({\rm mod} \;  3)$. 
Let us now consider the operator
\barray 
& H_X   =   - \sum_{n= \pm 1 }   \left( X_1^n X_2^{-n} X_5^n X_6^{-n} X_7^{-n} X_9^n 
\right.  &  \label{hx} \\
& +  \left.    X_2^n X_3^{-n} X_4^{-n} X_6^{n} X_7^{n} X_8^{-n}   + X_1^n X_3^{-n} X_4^{-n} X_5^{n} X_7^{-n} X_8^n  
\right)  & 
\nonumber 
\earray 
which satisfies 
\beq
[ H_Z, H_X] =0. 
\label{pa8}
\eeq
because the operators $X_i$ and $X^{-1}_j$   appear in  each triangle
and therefore their  product  commutes with  $Z_i Z_j Z_k$ (notice  that $Z X = q X Z)$. 
$H_X$  also commutes  with the charge $Q$, 
\beq
[ Q, H_X] =0, \quad Q= Z_1 Z_2 Z_3 \, , 
\label{pa9}
\eeq
which  implies that $H_Z$ and $H_X$ can be diagonalized simultaneously. 
To find the  ground state of $H$ we first minimize  $H_Z$ that  yields 
the states satisfying  the neutralization rule  Eq.(\ref{neu}) 
(recall Eq.(\ref{H})) 
\barray 
|H_{s_1, s_2, s_3}  \rangle & =   & |s_1, s_2, s_3, - s_1 - s_2, - s_2 - s_3 , -s_1 - s_3,  
\nonumber  \\
&  & |s_1 - s_2 + s_3, s_1 + s_2 - s_3, - s_1 + s_2 + s_3 \rangle . 
\nonumber 
\earray 
Now, we diagonalize $H_X$ in the subspace  with  $Q = e^{ 2 \pi i S}$,
\barray 
|H_{s_1, s_2, s_3}  \rangle_S  & =   & | H_{s_1, s_2, s_3}  \rangle, \quad s_1 + s_2 + s_3 = S , 
\nonumber 
\earray 
where it acts as 
\barray 
& H_X  |H_{s_1, s_2, s_3 }  \rangle_S  =    - \sum_{n=\pm 1} \left( 
|H_{s_1 + n, s_2 -n, s_3 }  \rangle_S \right.  & 
\nonumber  \\
& +  \left.   | H_{s_1, s_2 + n, s_3 - n }  \rangle_S + |H_{s_1+ n, s_2, s_3 -n}  \rangle_S  \right).   & 
\nonumber 
\earray 
This is a $9 \times 9$ matrix whose eigenvalues (degeneracies)  are :  $-6 (1), 3 (2), 0(6)$. The ground
state, i.e. $H_X = -6$,  is given by
\beq
|H \rangle_S = \frac{1}{3}  \sum_{s_1, s_2} | H_{s_1 , s_2 , S- s_1- s_2}  \rangle_S
\label{pa10}
\eeq
The uniqueness of this state and the fact that all its entries have the same sign 
follows from the Perron-Frobenius theorem applied to the non-positive matrix  $H_X$,  in the subspace $Q=e^{ 2 \pi i S}$.

How unique is the operator (\ref{hx})? To answer this question we shall consider 
the general expression
\beq
H_X = \sum_{n_1,\dots, n_9} C(\{ n_i \}) \prod_{i=1}^9 X_i^{n_i}, 
\label{pa11}
\eeq
and impose  the condition (\ref{pa8}) that yields 
\beq
q^{n_i + n_j + n_k } = 1 , \qquad i,j,k \in \tr \, , 
\label{pa12}
\eeq
which is solved by
\beq
n_i + n_j + n_k  = 0 \,  ({\rm mod} \, 3) , \qquad i,j,k \in \tr \,  . 
\label{pa13}
\eeq

This is a linear system of 9 equations for 9 unknowns. 
The rank of the corresponding matrix is 2. Choosing 0 in the RHS of Eq.(\ref{pa13})
and imposing translation invariance leads to the  ansatz (\ref{hx}). Finally, the sign of the 
constants $C(\{ n_i \})$ guarantees  that  (\ref{pa10}) are  GS's. 
Allowing the RHS of  Eq.(\ref{pa13}) to take the values 0 and  $\pm 3$ yields another solutions as for example  
\barray 
& H'_X    =   - \sum_{n= \pm 1 }   \left( X_1^n X_2^{n} X_4^n X_5^{-n} X_6^{-n} X_8^{-n}  
\right.  &  \label{hx2} \\
& +  \left.    X_2^n X_3^{n} X_4^{-n} X_5^{n} X_6^{-n} X_9^{-n}   + X_1^n X_3^{n} X_4^{-n} X_5^{-n} X_6^{n} X_7^{-n}   
\right)  & 
\nonumber 
\earray 
However,  this operator does not commute with the charge $Q$, 
so that the ground state of $H' = H_Z + H'_X$ is unique and given by
\beq
|H \rangle = \frac{1}{3^{3/2}}  \sum_{s_1, s_2, s_3} | H_{s_1 , s_2 , s_3 }  \rangle 
\label{pa14}
\eeq
A generalization of the Hamiltonian (\ref{hx}) to the torus $3^k \times 3^k$, with $k >1$
is given in Eq.(\ref{pa111}).  One can easily verify that this Hamiltonian satisfies
Eqs.(\ref{pa8}) and (\ref{pa9}), with $Q$ given in Eq.(\ref{constant}). The Hamiltonian (\ref{pa111}) 
contains $ 2 \times 3^{2 k -1}$ local operators $X_i^{\pm 1 }$, which for $k=2$ amounts to 54. 
However it is possible to construct Hamiltonians with a small number of terms, say 36 for $k=2$,
which satisfy Eqs.(\ref{pa8}) and (\ref{pa9}). Notice that 36 coincides with the Hamming dimension
for $k=2$, hence we expect the existence of parent Hamiltonians containing exactly $6^k$ local
terms $X_i^{\pm1}$. 

Finally, we want to remark that the neutralization rule can be extended to operators.
Indeed, let us consider  the following  operator defined on the boundary of the torus 
\beq
X_1^{n_1} X_2^{n_2} \dots X_N^{n_N}, \quad N = 3^k \, 
\label{pa15}
\eeq
which modifies the state on the boundary as 
\beq
|s_1, s_2 , \dots, s_N \rangle \rightarrow |s_1+ n_1, s_2 + n_2 , \dots, s_N + n_N \rangle
\label{pa16}
\eeq
The state on the second row can be obtained applying the neutralization rule to the RHS
of (\ref{pa16}), 
\beq
 |-s_1 - s_2 - n_1 -n_2,  - s_2 - s_3 - n_2 - n_3  , \dots  \rangle
\label{pa17}
\eeq
but this state can also be obtained acting on $|s_1, s_2 , \dots, s_N \rangle \rightarrow$
with the operator
\beq
X_1^{- n_1 - n_2} X_2^{- n_2 - n_3} \dots X_N^{- s_N - s_1}
\nonumber
\eeq
Proceeding in this manner one can associate an operator in the bulk to any boundary
operator (\ref{pa15}) on  the edge. In fact Eqs.(\ref{hx2}) and (\ref{hx2}) provide
examples of this holographic extension of operators.


\vspace{1cm} 


\end{document}